\documentclass{nseJournal}

\usepackage{bm}
\usepackage{times}
\usepackage[none]{hyphenat}
\usepackage{amsmath,amssymb}
\usepackage{fancyhdr}

\begin{document}

\renewcommand{\headrulewidth}{0.01pt}
\lhead{\emph{Accepted for publication in Fusion Science and Technology, published by Taylor \& Francis}}
\cfoot{}

\title{Deep Learning for the Analysis of Disruption Precursors based on Plasma Tomography}

\addAuthor{\correspondingAuthor{Diogo R. Ferreira}}{a}
\addAuthor{Pedro J. Carvalho}{a,c}
\addAuthor{Carlo Sozzi}{b}
\addAuthor{Peter J. Lomas}{c}
\addAuthor{JET Contributors}{$\dagger$}

\addAffiliation{}{EUROfusion Consortium, JET\\ Culham Science Centre\\ Abingdon, OX14 3DB, UK}

\addAffiliation{a}{Instituto de Plasmas e Fus\~{a}o Nuclear (IPFN)\\ Instituto Superior T\'{e}cnico (IST), Universidade de Lisboa\\ 1049-001 Lisboa, Portugal}

\addAffiliation{b}{Istituto per la Scienza e Tecnologia dei Plasmi (ISTP)\\ Consiglio Nazionale delle Ricerche (CNR)\\ 20125 Milano, Italia}

\addAffiliation{c}{Culham Centre for Fusion Energy (CCFE)\\ UK Atomic Energy Authority (UKAEA)\\ Abingdon, OX14 3DB, UK}

\addAffiliation{\dagger}{See the author list of E. Joffrin et al 2019 Nucl. Fusion 59 112021}

\correspondingEmail{diogo.ferreira@tecnico.ulisboa.pt}

\addKeyword{Plasma Tomography}
\addKeyword{Machine Learning}
\addKeyword{Anomaly Detection}

\titlePage

\begin{abstract}
The JET baseline scenario is being developed to achieve high fusion performance and sustained fusion power. However, with higher plasma current and higher input power, an increase in pulse disruptivity is being observed. Although there is a wide range of possible disruption causes, the present disruptions seem to be closely related to radiative phenomena such as impurity accumulation, core radiation, and radiative collapse. In this work, we focus on bolometer tomography to reconstruct the plasma radiation profile and, on top of it, we apply anomaly detection to identify the radiation patterns that precede major disruptions. The approach makes extensive use of machine learning. First, we train a surrogate model for plasma tomography based on matrix multiplication, which provides a fast method to compute the plasma radiation profiles across the full extent of any given pulse. Then, we train a variational autoencoder to reproduce the radiation profiles by encoding them into a latent distribution and subsequently decoding them. As an anomaly detector, the variational autoencoder struggles to reproduce unusual behaviors, which includes not only the actual disruptions but their precursors as well. These precursors are identified based on an analysis of the anomaly score across all baseline pulses in two recent campaigns at JET.
\end{abstract}

\section{Introduction}

The JET baseline scenario~\cite{garzotti19scenario} is being developed to establish the best recipe for achieving high fusion performance and sustained fusion power with a view towards Deuterium-Tritium (D-T) operation in the near future. However, as the plasma current and heating power are being increased, a higher rate of pulse disruptivity is also being observed. The need to bring disruptivity under control is currently one of the most pressing issues in scenario development, especially when considering that JET experiments are intended as a testbed for ITER-relevant plasma scenarios~\cite{donne16risk}.

Whether disruptivity can be brought down by avoiding the paths that lead to disruption, or by taking mitigating actions whenever the experiment is found to be in one of such paths, is still an open question, and in fact both options can be pursued independently~\cite{sozzi18early}. In any case, identifying and characterizing disruptive behaviors is of paramount importance not only to understand the underlying physical phenomena, but also to make the best decisions in the context of experiment design and plasma control.

In the past, a landmark paper~\cite{vries11survey} identified a wide range of disruption causes at JET, including multiple physical phenomena as well as a variety of technical issues; the study was conducted when JET still had a carbon wall. More recently, after the introduction of the ITER-like wall~\cite{matthews13ilw}, it was found that the present disruptions are closely related to radiative phenomena~\cite{joffrin13first}, by a sequence of events that includes impurity transport from the edge to the core, impurity accumulation at the core, development of strong core radiation, and radiative collapse when the radiated power exceeds the input power.

It is therefore important to analyze the plasma radiation profile in search for any symptoms that may represent disruption precursors. One of such symptoms is strong core radiation, but there may be others. The main goal of this work is to find disruption precursors by automated means, namely by using machine learning to identify anomalous plasma profiles that precede disruptions in a time frame that is relevant for disruption avoidance and mitigation, and which may or may not have been considered before.

In principle, this could be framed as a pure anomaly detection problem, if it were not for the fact that reconstructing the plasma radiation profile is a significant computational problem \emph{per se}. At JET, the plasma radiation profile is reconstructed from the bolometer diagnostic~\cite{huber07upgraded} with tomography techniques~\cite{ingesson08tomography}. For the purpose of anomaly detection, it will be necessary to scan through millions of plasma profiles (to be quantified more precisely later). Although there has been progress in developing faster tomography algorithms~\cite{loffelmann16tomography}, deep learning provides a viable alternative through the use of neural networks that can compute thousands of reconstructions per second when running on a Graphics Process Unit (GPU)~\cite{ferreira18fullpulse}.

Deep learning is also a key element in the anomaly detection task, through the use of a variational autoencoder (VAE)~\cite{kingma14vae} to detect unusual patterns in the plasma radiation profiles. In the literature, VAEs have been used to perform anomaly detection on different kinds of data, including time series~\cite{pereira18unsupervised}, medical images~\cite{chen18unsupervised} and graph data~\cite{kipf16variational}. In this work, a VAE provides an anomaly score on each plasma radiation profile, and by computing that anomaly score across an entire pulse, one can obtain a helpful indication of where the disruption precursors can be found.

The following sections describe the approach in more detail, together with a selection of representative results. In particular, Sec.~\ref{sec_bolometry} introduces the bolometer diagnostic at JET and its associated tomography techniques; Sec.~\ref{sec_approach} describes the deep learning models used in this work, for tomographic reconstruction and anomaly detection, respectively; Sec.~\ref{sec_results} discusses the results obtained with those models on sample pulses that are representative of the baseline scenario; finally, Sec.~\ref{sec_conclusion} concludes the paper.

\section{Bolometer tomography at JET}
\label{sec_bolometry}

The bolometer diagnostic at JET~\cite{huber07upgraded} comprises two cameras -- a horizontal camera and a vertical camera -- with 24 bolometers each. In essence, a bolometer consists of a thin metal foil (about 10 µm) coupled with a
temperature-sensitive resistor. As the metal foil absorbs radiation power, its temperature changes and there is a proportional change in resistance, which can be measured to provide linear response to radiation in the ultraviolet to soft X-ray range~\cite{mast91bolometer}.

\begin{figure}[h]
	\centering
	\includegraphics[scale=0.35]{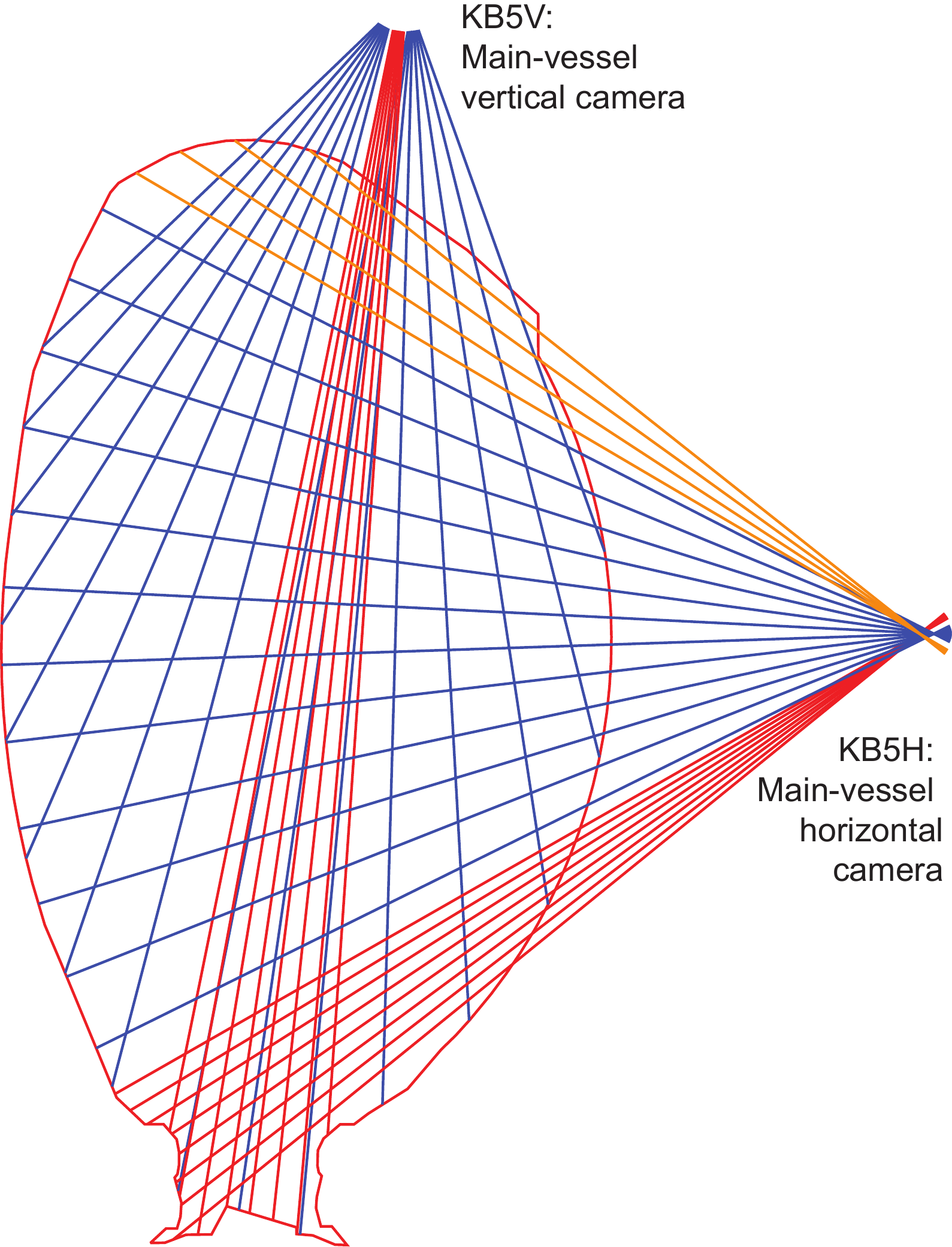}
	\caption{Lines of sight for the bolometer diagnostic at JET.}
	\label{fig_kb5}
\end{figure}

Each bolometer measures the line-integrated radiation along a particular line of sight. For the horizontal camera, the lines of sight are defined by a pinhole structure, whereas for the vertical camera they are defined by a collimator block~\cite{mccormick05bolometry}. The geometrical arrangement of these lines of sight is such that, for each camera, 16 channels cover the whole plasma, and 8 channels provide a more fine-grained resolution over the divertor region. This arrangement is illustrated in Fig.~\ref{fig_kb5}.

From the bolometer measurements, there are several different tomography techniques that can be applied to reconstruct the plasma radiation profile~\cite{mlynar19current}. The method that is used at JET uses an
iterative constrained optimization algorithm that minimizes the error with respect to the bolometer measurements, while
requiring the solution to be non-negative~\cite{ingesson98tomography}. This iterative procedure takes a significant amount of time, typically on the order of minutes to produce a single reconstruction.

Recently, it was shown that it is possible to produce essentially the same results with a deep neural network, at the cost of a small amount of error, but several orders of magnitude faster~\cite{ferreira18fullpulse}. In this case, the network has a structure that resembles the inverse of a convolutional neural network (CNN), with two dense layers to preprocess the one-dimensional bolometer measurements, followed by a number of transposed convolutions to output a two-dimensional plasma radiation profile.

\begin{figure}[h]
	\centering
	\includegraphics[width=\textwidth]{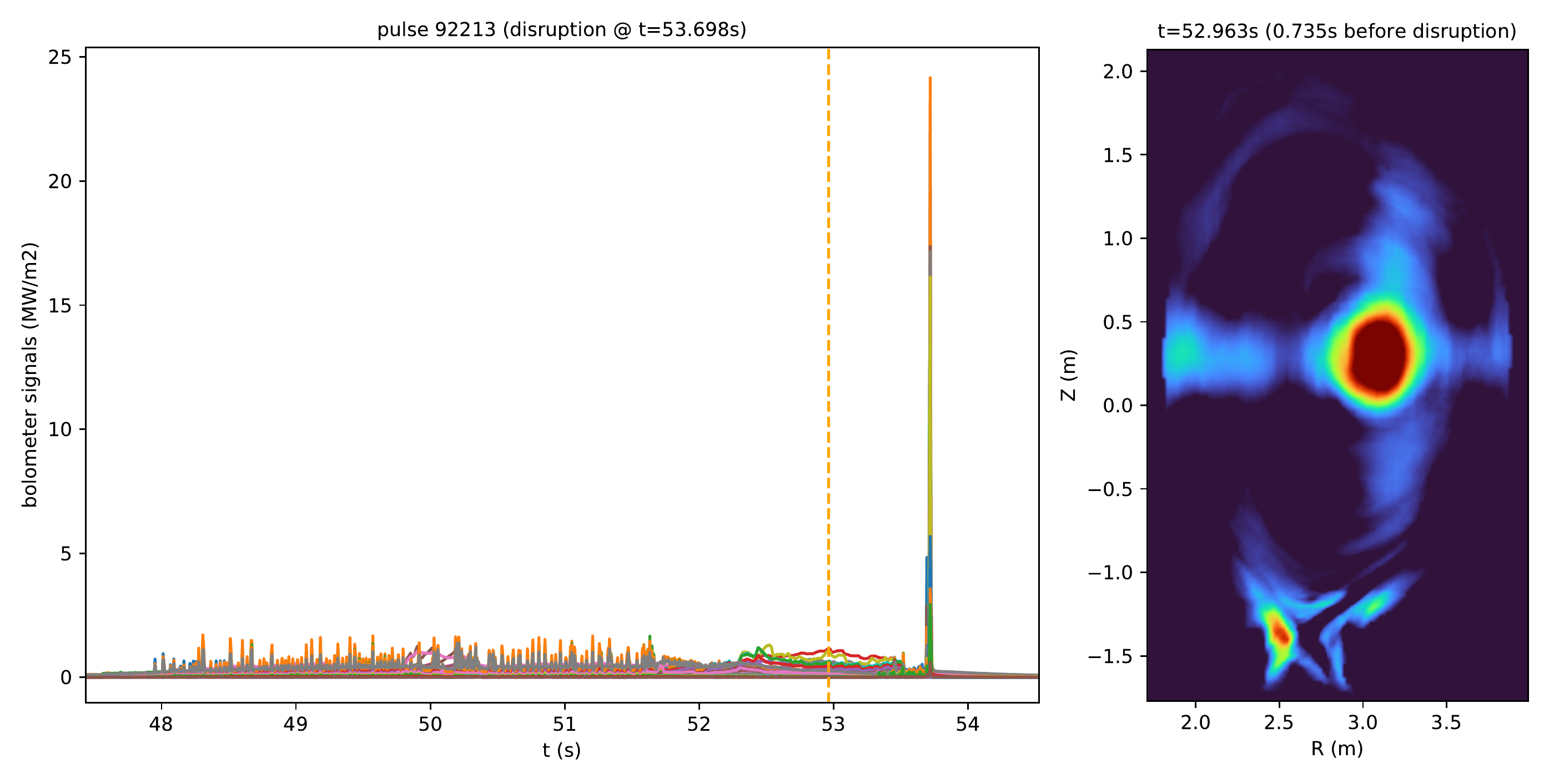}
	\caption{Bolometer signals for pulse 92213 (\emph{left}) and tomographic reconstruction of the plasma radiation profile at about $t$=53.0s (\emph{right}). The pulse disrupts at $t$=53.7s.}
	\label{fig_bolo_signals}
\end{figure}

Fig.~\ref{fig_bolo_signals} shows an example of a plasma radiation profile computed at a particularly interesting point of a disruptive pulse. This is a point in time that is still reasonably far from the disruption and where the bolometer signals display a distinct behavior from the rest of the pulse. At this point, the strongest signals come from the bolometers that are pointing at the plasma core and, as can be observed in the plasma profile, there is a strong focus of radiation at the core that is a sign of impurity accumulation in that region. This is corroborated by other diagnostics, which show density peaking and a hollow temperature profile.

Since the geometry of the lines of sight is fixed, some authors have used a single camera (e.g.~the horizontal one) or even a subset of those bolometers to detect core radiation without performing a full reconstruction of the plasma radiation profile~\cite{pau18indicators}. However, this provides only a simplified view, because the horizontal camera measures the spread of radiation along the vertical axis, but the vertical camera is needed in order to resolve the spread of radiation along the horizontal axis.

In this work, rather than analyzing disruption precursors based on the activity of certain bolometers, we would like to characterize those precursors based on the actual shapes or patterns that appear in the plasma radiation profile. This will always require reconstructing the plasma profile from the bolometer signals, so we will need to devise an expeditious method to do that.

In addition, we purposefully intend to make \emph{tabula rasa} of any preconceptions about how the disruption precursors look like, in order to check whether the precursors that we find agree with our own preconceptions. For example, if it is true that the development of strong core radiation is an important disruption precursor, then this should eventually stand out in the results of our analysis.

\section{Anomaly detection approach}
\label{sec_approach}

Fig.~\ref{fig_bolo_signals} showed the time evolution of the bolometer signals across a sample pulse. But just as there is a time evolution of the bolometer signals, there is a corresponding time evolution of the plasma radiation profile. The difference is that the bolometer signals are provided directly by the diagnostic, whereas the plasma profile must be reconstructed by a computational process over those signals.

A proper reconstruction of the plasma radiation profile will have to perform the tomographic inversion of the bolometer measurements, and also smooth out fluctuations due to random noise and malfunctioning detectors. This explains why performing anomaly detection on top of the plasma profiles is more promising than performing it directly on the bolometer signals: because a layer of preprocessing has already been applied over the raw signals. Naturally, neural networks excel at such preprocessing, because they assign different weights to each input and also apply convolutions to smooth out the results~\cite{ferreira18fullpulse}.

Once the radiation profiles have been computed, a different kind of deep learning model can be employed to detect anomalies. In particular, variational autoencoders (VAEs)~\cite{kingma14vae} have been extensively used for anomaly detection in a number of different applications~\cite{pereira18unsupervised,chen18unsupervised,kipf16variational}. When applied to plasma radiation profiles, a VAE tries to reproduce whatever profile is given as input. Because the VAE is inherently stochastic and is exposed to a much larger fraction of typical behaviors than atypical ones, it will learn to reproduce the former much better than the latter.

Hence, when given an anomalous profile as input (which, in our case, may be a disruption precursor or some other odd event during the pulse), the VAE will reproduce it with a larger reconstruction error than it would otherwise do for a regular profile. This reconstruction error becomes an anomaly score for the profile that is provided as input. An analysis of the anomaly score across an entire pulse will point to unusual behaviors of the kind that we are interested in finding.

\begin{figure}[h]
	\centering
	\includegraphics[width=\textwidth]{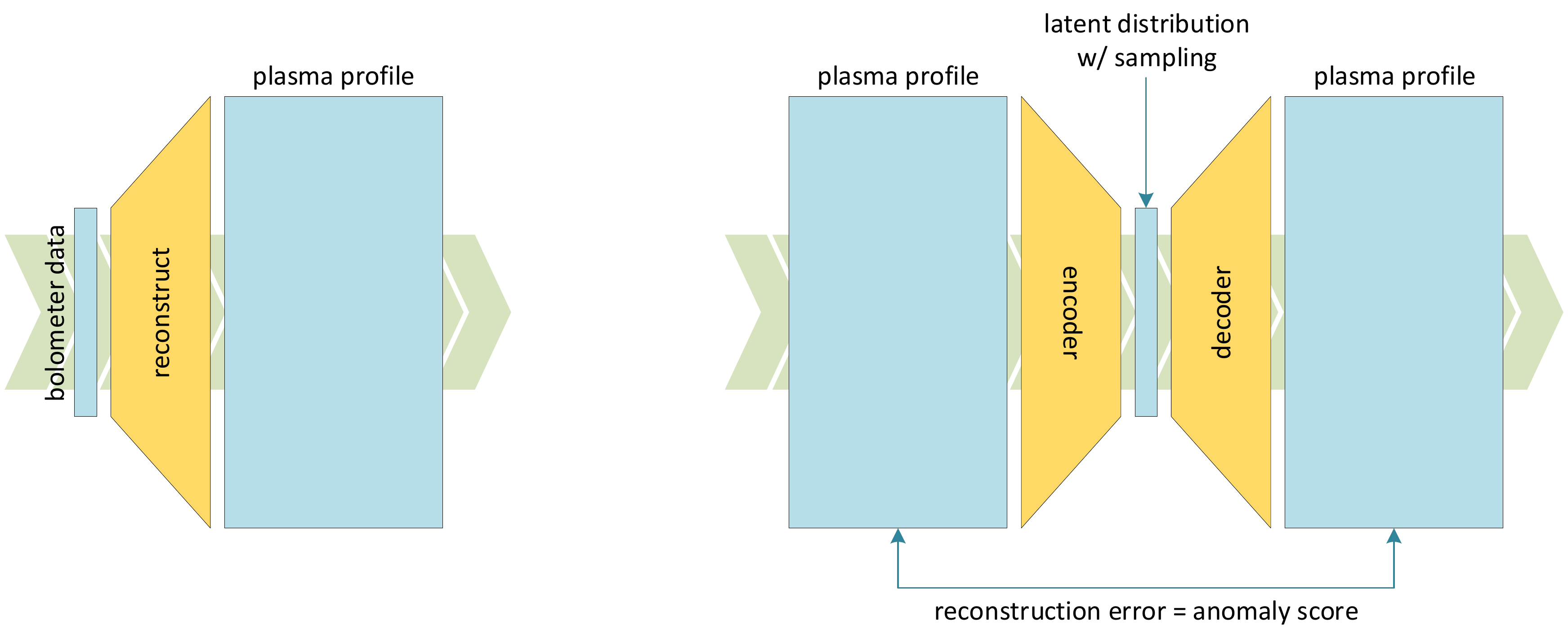}
	\caption{Deep learning models used in this work: tomographic reconstruction model (\emph{left}) and variational autoencoder for anomaly detection (\emph{right}).}
	\label{fig_models}
\end{figure}

In summary, to apply anomaly detection on top of the radiation profile, two different components are necessary: (1) a fast tomographic method to generate radiation profiles from bolometer data, e.g.~a neural network or even a simpler model; and (2) an anomaly detector to point out unusual profiles, i.e.~a variational autoencoder. Fig.~\ref{fig_models} illustrates both models and the next subsections delve into the details.

\subsection{Tomographic reconstruction model}

A question that arises when looking at the models in Fig.~\ref{fig_models} is whether they must remain separate or can be combined into a single model. For training purposes, they should remain separate, because the tomographic reconstruction model is trained on pairs of input-output samples (a sample of bolometer data together with the corresponding plasma profile that the model should approximate), whereas the anomaly detection model is trained on input samples only (the plasma profile that the model will try to reproduce at the output). This is essentially the difference between supervised and unsupervised learning.

However, once the tomographic reconstruction model is trained, a possible approach is to use it as a building block in the anomaly detection model, an idea that is akin to the use of pre-trained models in deep learning~\cite{marmanis16pretrained}. In this case, the pre-trained tomographic reconstruction model would be inserted as a first layer into the anomaly detection model, and the variational autoencoder would be trained from a bolometer data input (although the reconstruction loss would still be based on plasma profiles).

Such approach is only feasible if the tomographic reconstruction model is relatively lightweight. Otherwise, the anomaly detection model, with the addition of the tomographic reconstruction model, becomes too complex and computationally expensive to train. (For the interested reader, the first bottleneck is that such model would take up a large fraction of GPU memory, and therefore would have to be trained with a small batch size, leading up to a lengthy and noisy training process.)

On the other hand, the possibility of keeping the models separate is not without its drawbacks as well. In this case, we can afford to have a more complex tomographic reconstruction model, such as a deconvolutional neural network with multiple layers. However, once the model is trained, we will need to pre-compute all the plasma radiation profiles that will be used to train the variational autoencoder. Although this might not take much time when using a GPU, it will certainly take a lot of space, and when it comes to training the variational autoencoder, there will be a lot of data shuffling between CPU and GPU, again leading up to a sub-optimized and under-performing training process.

The solution we adopted to deal with this problem was to use a lightweight model. Instead of using a deep neural network as in~\cite{ferreira18fullpulse}, we reduced the tomographic reconstruction model to a single layer that computes the plasma radiation profile by matrix multiplication over the bolometer data.

More precisely:
\begin{itemize}

\item Let $\bm{X}=\{X_{ij}\}$ be an $N\!\times\!48$ array representing a batch of bolometer data, where $N$ is an arbitrary batch size and $48$ is the number of lines of sight.

\item Let $\bm{M}=\{M_{jkm}\}$ be a $48\!\times\!196\!\times\!115$ array representing the model parameters, where $196\!\times\!115$ corresponds to the pixel resolution of the plasma radiation profile.

\item Then $\bm{Y}=\{Y_{ikm}\}$ with $Y_{ikm}\!=\!\sum_{j} X_{ij} M_{jkm}$ is an $N\!\times\!196\!\times\!115$ array representing a batch of plasma profiles reconstructed from the bolometer data.

\item Given a batch of training data $(\bm{\widetilde{X}},\bm{\widetilde{Y}}) = (\{\widetilde{X}_{ij}\},\{\widetilde{Y}_{ikm}\})$, the model is trained by gradient descent to minimize the mean absolute error:
\begin{equation}
L(\bm{M}) = \frac{1}{N\!\times\!196\!\times\!115} \sum_{i=1}^{N} \sum_{k=1}^{196} \sum_{m=1}^{115} \lvert \widetilde{Y}_{ikm} - \sum_{j=1}^{48} \widetilde{X}_{ij} M_{jkm} \rvert
\label{eq_tomo_loss}
\end{equation}

\end{itemize}

In essence, the model is a $48\!\times\!196\!\times\!115$ array, where each of its $196\!\times\!115$ outputs is generated by a weighted sum of its 48 inputs, and the weights are learned by minimizing the expression in Eq.~(\ref{eq_tomo_loss}).

An interesting feature of the model above is that the batch size $N$ can be made arbitrarily large (within the contraints of available memory). In particular, $N$ can be made as large as the length of an entire pulse, and in this case the plasma profiles for an entire pulse can be computed with a single matrix multiplication, either on CPU or on GPU. Moreover, if the GPU memory allows it, $N$ can be made as large as the length of the training dataset, and in this case the full dataset is loaded into GPU memory and the model can be trained very efficiently, with no data transfers between CPU and GPU.

\begin{figure}[h]
	\centering
	\includegraphics[scale=0.55]{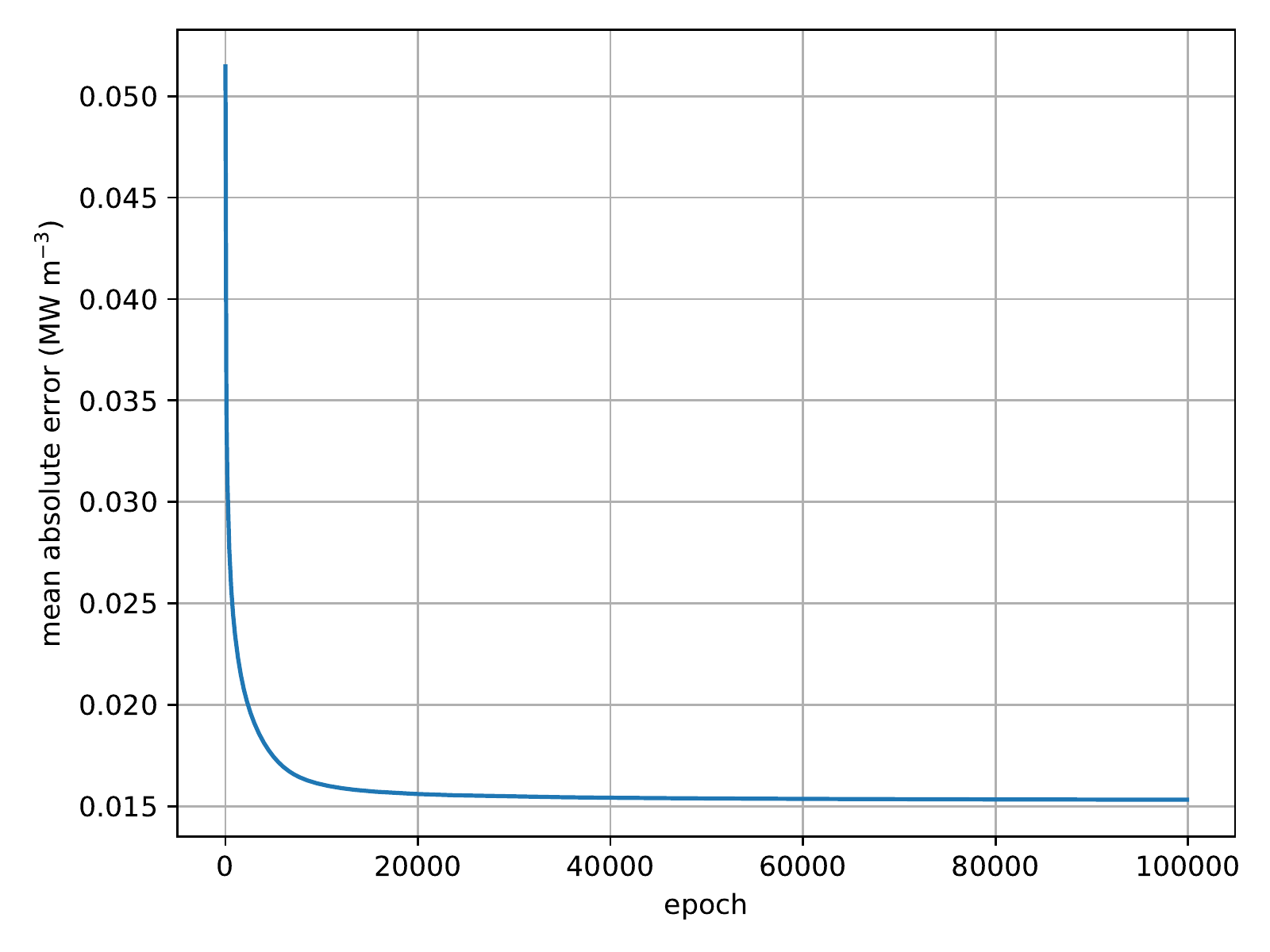}
	\caption{Loss during training the tomographic reconstruction model.}
	\label{fig_train_tomo}
\end{figure}

For training the model, we collected a dataset of about 10 000 samples profiles, where we exercised extra care to select high-quality reconstructions while avoiding profiles with noticeable artifacts. Since we are training a linear model on a number of samples that is much larger than the number of inputs, there is no danger of overfitting, so the model was trained on the entire dataset. Fig.~\ref{fig_train_tomo} shows the evolution of the training loss given by Eq.~(\ref{eq_tomo_loss}) across 100 000 epochs. With all data loaded into GPU memory (i.e.~one single training batch), this took about one hour to run on an Nvidia Tesla P100 GPU.

In terms of accuracy in reconstructing the plasma radiation profile, this lightweight model achieves a mean absolute error close to 0.015 MW/m$^\text{3}$, while the deep neural network in~\cite{ferreira18fullpulse} has a reported error of 0.010 MW/m$^\text{3}$. Although the training dataset is not the same, this provides a sense of how much error we incur by simplifying the model. Naturally, the deep neural network has better accuracy since it comprises a series of non-linear layers, whereas the model that we use here is equivalent to a single linear layer.

\begin{figure}[h]
	\centering
	\includegraphics[width=\textwidth]{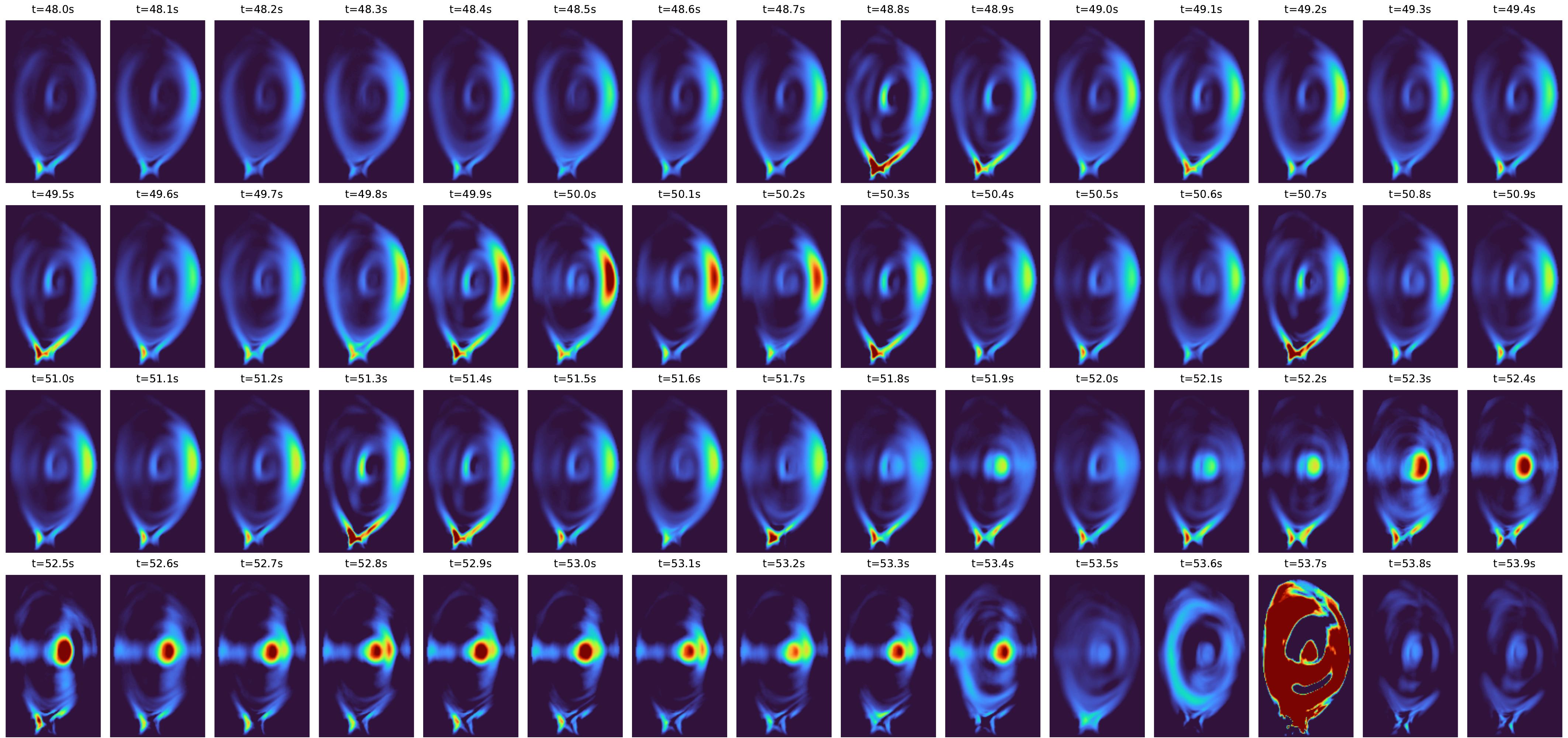}
	\caption{Plasma radiation profiles for pulse 92213 from t=48.0s to t=53.9s with a time step of 0.1s.}
	\label{fig_full_pulse}
\end{figure}

Fig.~\ref{fig_full_pulse} shows an example of the plasma radiation profiles produced by the model. Despite its simplicity, all the usual features, such as the development of radiation blobs at the outer edge and at the plasma core, as well as divertor activity, are clearly recognizable.

\subsection{Anomaly detection model}

The anomaly detection model works on the basis of the plasma profiles provided by the tomographic reconstruction model. As illustrated earlier in Fig.~\ref{fig_models}, what it tries to do is to reproduce the plasma radiation profile through the mechanisms of a variational autoencoder.

While any neural network that produces an output of the same type and shape as its input can considered to be an autoencoder, the variational autoencoder has the distinctive feature of encoding the input data as a probability distribution (the so-called latent distribution), and then generating the output by sampling from that distribution. As a result, the output is a stochastic approximation of the input. This approximation will be accurate for the training data that the model has already seen; it is only for data that the model has not seen before that the reconstruction error will be large.

Rather than allowing the parameters of the latent distribution to be learned freely, which could cause the VAE to overfit the training data and reduce its stochasticity to zero, it is common to impose a prior on the latent distribution. A typical choice is a standard multivariate normal $\mathcal{N}(\bm{0},\bm{I})$ of the same dimension as the latent distribution. This can be seen as a form of regularization over the latent space. With this regularization, the VAE is trained to minimize both the reconstruction error and the Kullback–Leibler (KL) divergence of the latent distribution with respect to the prior distribution~\cite{kingma19introduction}.

In our case, the plasma profile is a two-dimensional structure which can be processed as an image. Therefore, we use a series of convolutional layers to perform the encoding into the latent space. This latent space is the parameter space of a multivariate normal $\mathcal{N}(\bm{\mu},\bm{\Sigma})$ characterized by its mean vector $\bm{\mu}$ and covariance matrix $\bm{\Sigma}$. For a $d$-dimensional distribution, $\bm{\mu}$ has length $d$ and $\bm{\Sigma}$ is a $d\!\times\!d$ symmetric matrix. Due to this symmetry, we need to specify only $d\!+\!d(d\!+\!1)/2$ parameters.

To generate an output profile, the VAE draws a sample from the $d$-dimensional latent distribution, and then applies a series of transposed convolutions to bring it up to the desired size and shape.

\begin{figure}[h]
	\centering
	\includegraphics[width=\textwidth]{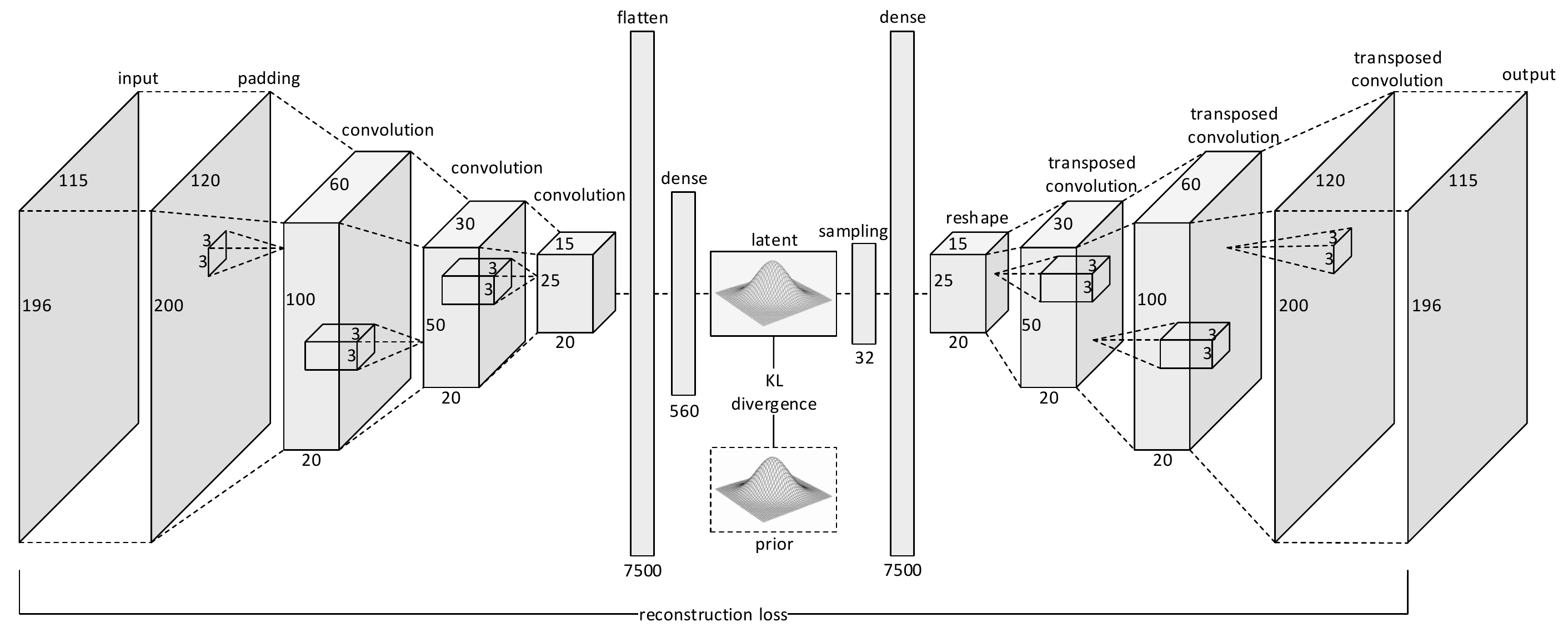}
	\caption{Arquitecture of the variational autoencoder for anomaly detection.}
	\label{fig_vae}
\end{figure}

Fig~\ref{fig_vae} shows the internal structure of the VAE, where we use a latent space of $d\!=\!32$. The rationale for this choice is that the plasma profile is obtained from a 48-channel diagnostic where there is some redundancy (due to correlation between channels) and also a few malfunctioning detectors, so the amount of information needed to reconstruct the plasma profile can be a bit smaller. With a choice of $d\!=\!32$, the numbers of parameters for the latent distribution is $d\!+\!d(d\!+\!1)/2=560$.

To train this VAE, we collected the bolometer data for about 400 baseline pulses from two recent campaigns at JET (Nov~2015 -- Mar~2020). As is common when doing bolometer tomography at JET, we downsampled the bolometer signals with a 5-ms window average for noise reduction, yielding an effective sampling rate of 200 Hz. Since each pulse typically lasts for around 30s, the total number of samples is about $\text{200}\!\times\!\text{30}\!\times\!\text{400}\!\sim\!\text{2.4}$ million. In practice, we got a number close to 2 million.

By passing these bolometer data through the tomographic reconstruction model, we obtain the plasma profiles needed to train the VAE. However, rather than computing them before training, the profiles are computed on-the-fly, during training, by incorporating the tomographic reconstruction model as a first layer in the VAE. Naturally, the weights of the tomographic reconstruction model are frozen, so that they do not change as the VAE is being trained. This way, the VAE is trained directly from bolometer data.

Another important aspect is that the VAE is trained on bolometer data from non-disruptive pulses only, so as to keep the anomalies of disruptive pulses out of the training data, since these will be reserved for subsequent analysis. This is an important difference between anomaly detection and disruption prediction models, as the latter are usually trained and tested on subsets that include both disruptive and non-disruptive pulses~\cite{moreno16prediction}. In our case, we train the VAE on non-disruptive pulses only, and test it on disruptive pulses only, so the usual metrics used for binary classification do not apply here (for example, there are no false alarms since we do not test on negative examples). The VAE is fundamentally an analysis support tool to identify disruption precursors that stand out as unusual profiles in disruptive pulses.

Since the ratio of non-disruptive to disruptive pulses in the baseline scenario is about 60\%/40\%, one could think that the VAE would be trained on about 60\% of data and tested on the remaining 40\%. However, disruptive pulses are usually shorter, so in terms of number of plasma profiles the ratio is more like 70\%/30\%. These 70\% correspond to about 1.4 million training profiles.

\begin{figure}[h]
	\centering
	\includegraphics[scale=0.55]{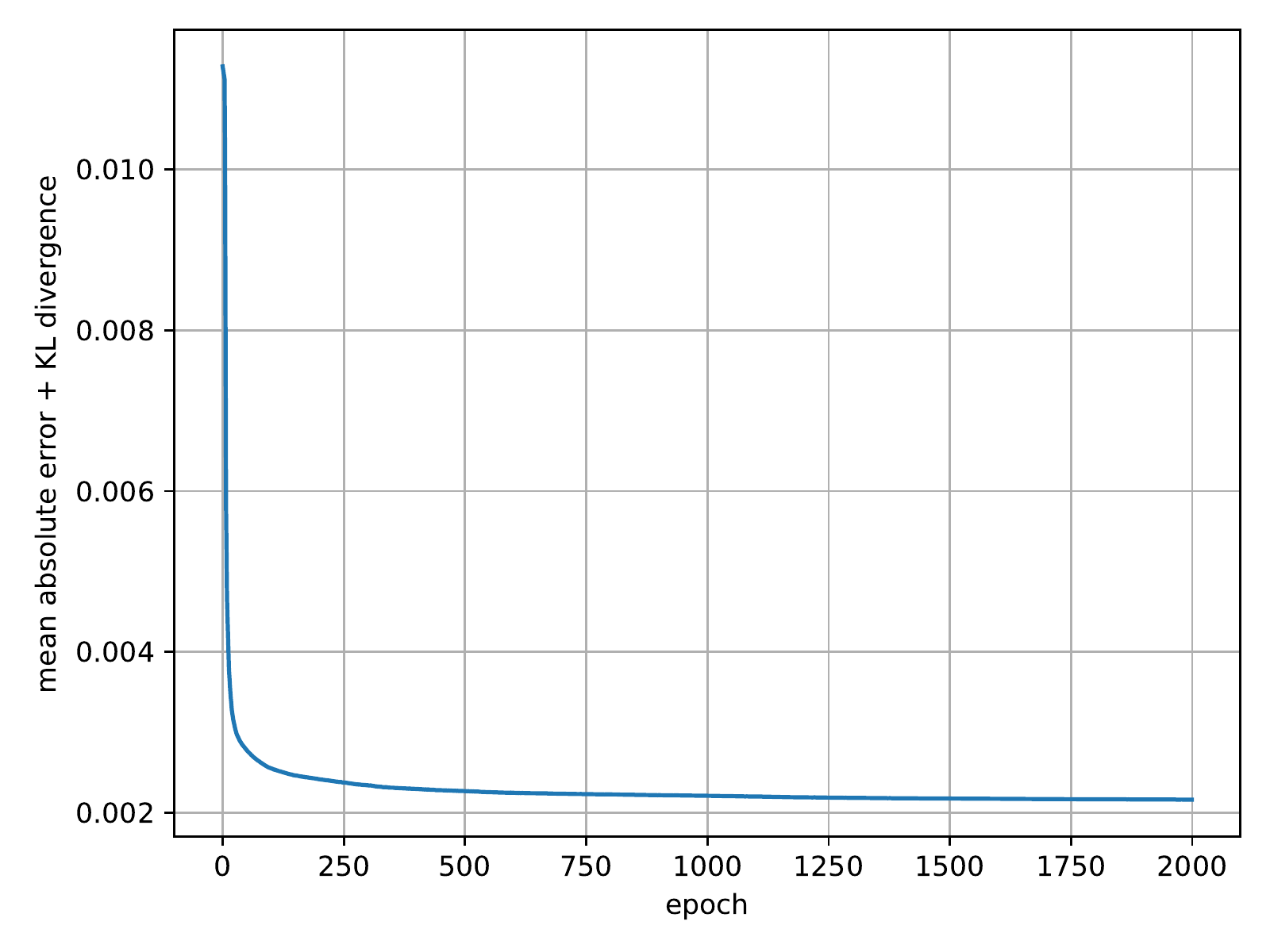}
	\caption{Loss during training of the anomaly detection model.}
	\label{fig_train_bolo}
\end{figure}

Fig.~\ref{fig_train_bolo} shows the evolution of the training loss across 2000 epochs with 50 batches (or updates) per epoch, which ran for 15 hours on 8 GPUs. The actual value of the training loss is somewhat meaningless as it includes the contribution of two terms -- the mean absolute error of the reconstruction, plus the KL divergence of the latent distribution with respect to the prior -- where the weight of the second term can be adjusted to enforce a stronger or weaker regularization.

In any case, Fig.~\ref{fig_train_bolo} illustrates the convergence of the training process and also the fact that, since the KL divergence is always non-negative, by comparing Fig.~\ref{fig_train_bolo} to Fig.~\ref{fig_train_tomo} one concludes that the VAE is better at reproducing the plasma profile than the tomographic reconstruction model is at reconstructing the plasma profile from the original bolometer data. To some extent this is not surprising, since the VAE has more learning capacity and has also been given an easier task.

\section{Analysis of disruption precursors}
\label{sec_results}

Having trained the anomaly detection model on non-disruptive pulses, we now apply it on disruptive pulses to analyze the anomaly score. For this purpose, the anomaly score is the reconstruction loss of the VAE at each point in time, as given by the mean absolute error between the input and output profiles. The regularization term based on the KL divergence is no longer included, as it applies during training only.

\begin{figure}[p]
	\centering
	\includegraphics[width=\textwidth]{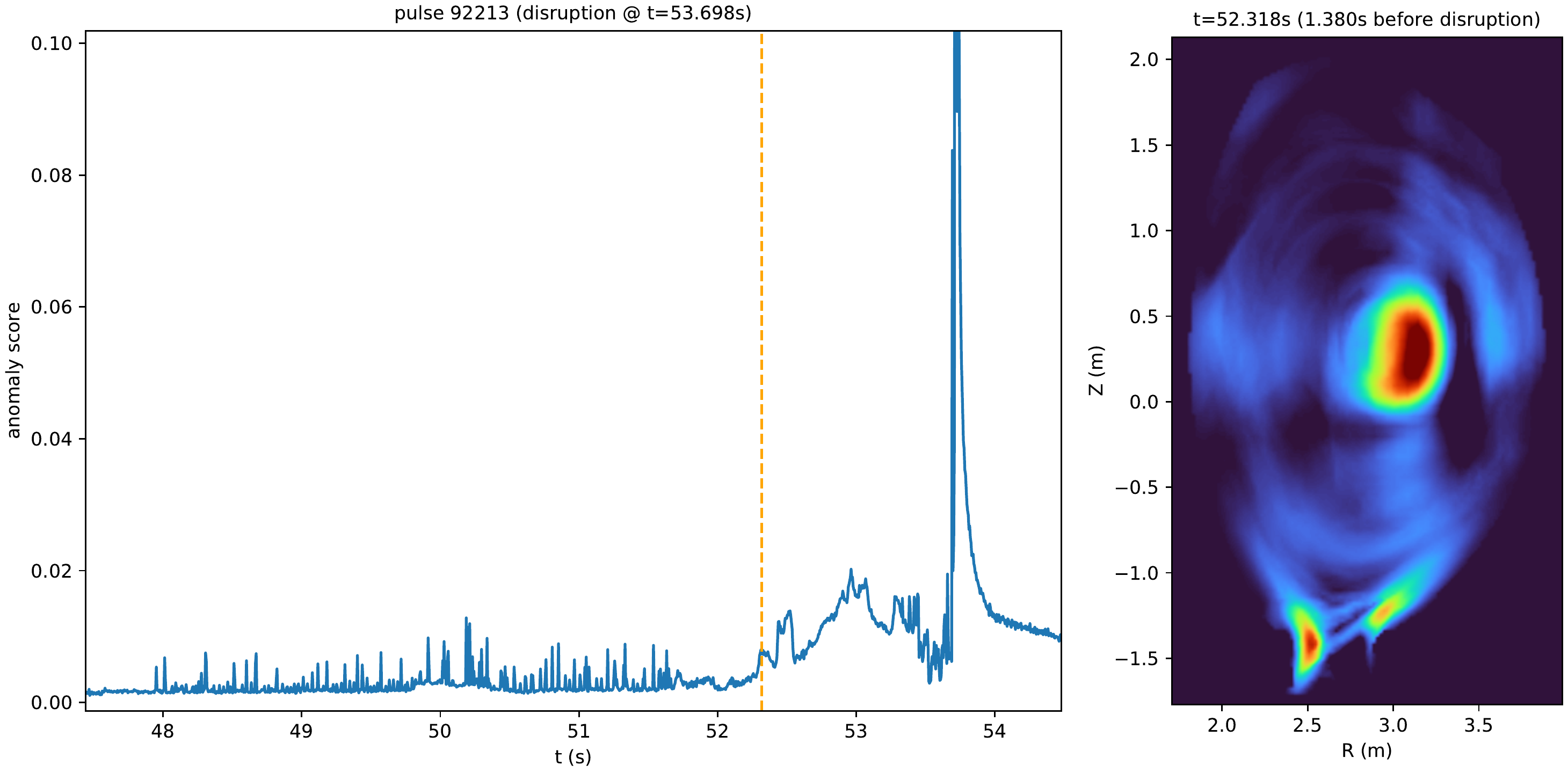}
	\caption{Anomaly score for pulse 92213 (\emph{left}) and tomographic reconstruction at a point when core radiation begins to develop (\emph{right}).}
	\label{fig_anomaly_1}
\end{figure}

Fig.~\ref{fig_anomaly_1} shows the anomaly score provided by the VAE for the same pulse that was used as an example before. As can be seen in the figure, there is a point at which the anomaly score begins to rise and stays relatively high for an extended period of time before the disruption. This corresponds to the moment when strong core radiation begins to develop and a radiation blob settles into the core.

\begin{figure}[p]
	\centering
	\includegraphics[width=\textwidth]{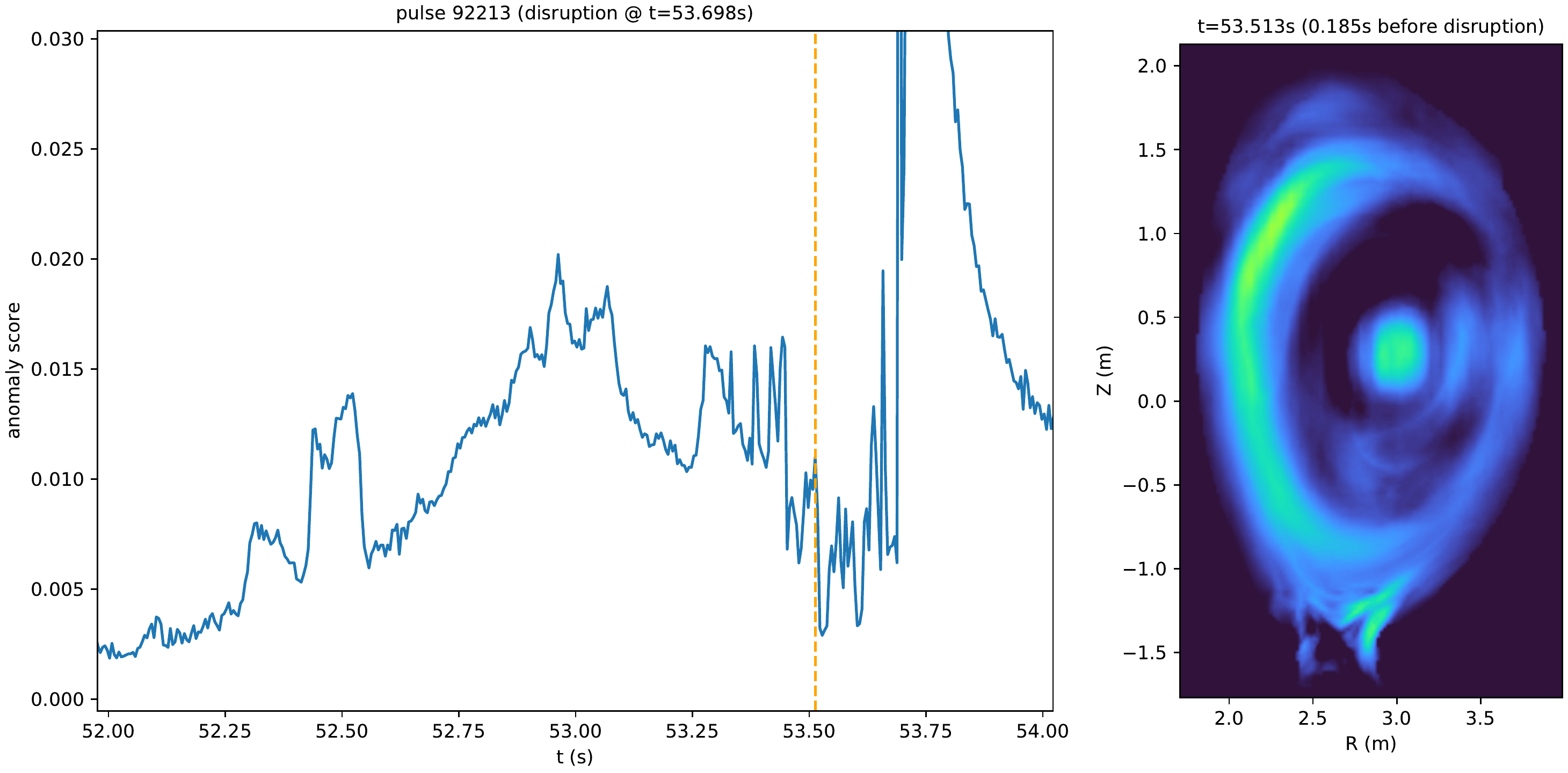}
	\caption{Anomaly score for pulse 92213 (\emph{left}) and tomographic reconstruction at a point when a burst of radiation appears on the high-field side (\emph{right}).}
	\label{fig_anomaly_2}
\end{figure}

But the anomaly score highlights something else too. At about $t$=53.45s there is a sudden drop in core radiation, which does not return to its former levels, and a different kind of instability appears. As illustrated in Fig.~\ref{fig_anomaly_2}, the plasma starts displaying a pulsating behavior characterized by periodic bursts of radiation on the high-field side, which grow more intense towards the disruption. This behavior is consistent with the well-known phenomena of MARFE (multifaceted asymmetric radiation from the edge)~\cite{lipschultz84marfe}.

Indeed, an analysis of other plasma diagnostics reveals that these appear to be minor disruptions driven by MARFE behavior, with spikes in radiated power being accompanied by several other symptoms, such as spikes in impurity influx, small but noticeable blips in the plasma current, and an increase in the locked mode amplitude, which is one of the main disruption predictors used at JET and other tokamaks~\cite{vega15predictor,aledda15improvements,ding13prediction}. In contrast, the temperature profile becomes hollow and the density profile becomes peaked much earlier, when core radiation starts to grow at around $t$=52.3s.

These two time frames illustrate the difference between disruption avoidance and mitigation. When core radiation starts developing at $t$=52.3s, we are still far away from the disruption; sufficiently far, in fact, to change the course of the experiment. For example, by applying ICRH (ion cyclotron resonance heating), it is possible to recover from a hollow temperature profile and create the conditions for a safe termination~\cite{lerche16icrh}. On the other hand, when the MARFE behavior shows up at $t$=53.5s, it is now too late for avoidance; a locked mode has already settled in, and the only option is to take mitigating actions, namely by triggering the DMV (disruption mitigation valve)~\cite{reux13dmv}.

\begin{figure}[h]
	\centering
	\includegraphics[width=\textwidth]{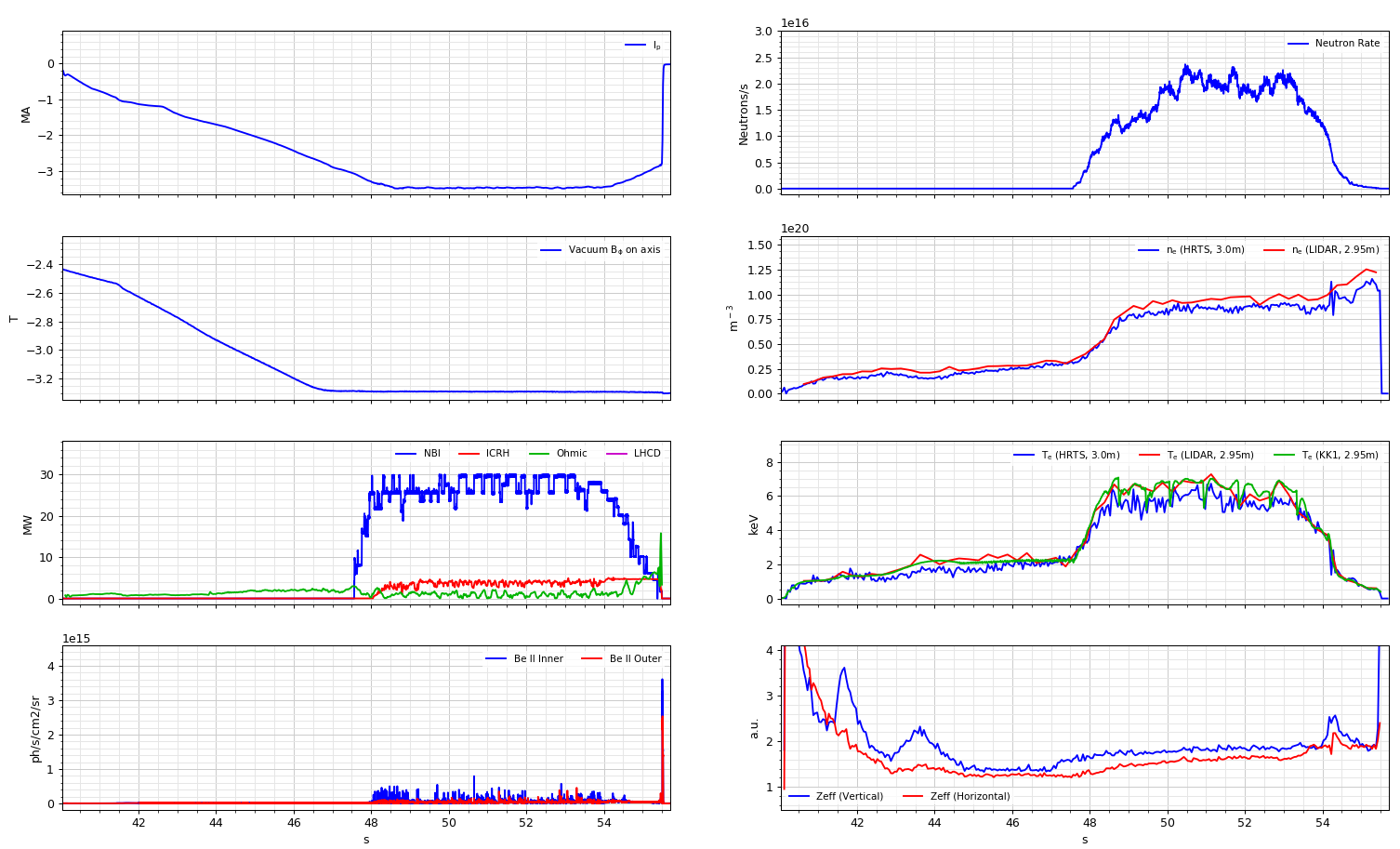}
	\caption{Main characteristics of pulse 96486, including plasma current, magnetic field, heating power, impurity influx (\emph{left}), and neutron yield, electron density, electron temperature, and effective charge (\emph{right}).}
	\label{fig_summary}
\end{figure}

As a second example, we now turn our attention to a pulse that achieved a record neutron yield at the time of the experiment (Dec~2019). Figure~\ref{fig_summary} shows the main characteristics of this pulse, where it is possible to observe that the pulse goes to the full designed length and starts the planned ramp down of power at $t$=54.0s, but it disrupts at around $t$=55.5s.

\begin{figure}[p]
	\centering
	\includegraphics[width=\textwidth]{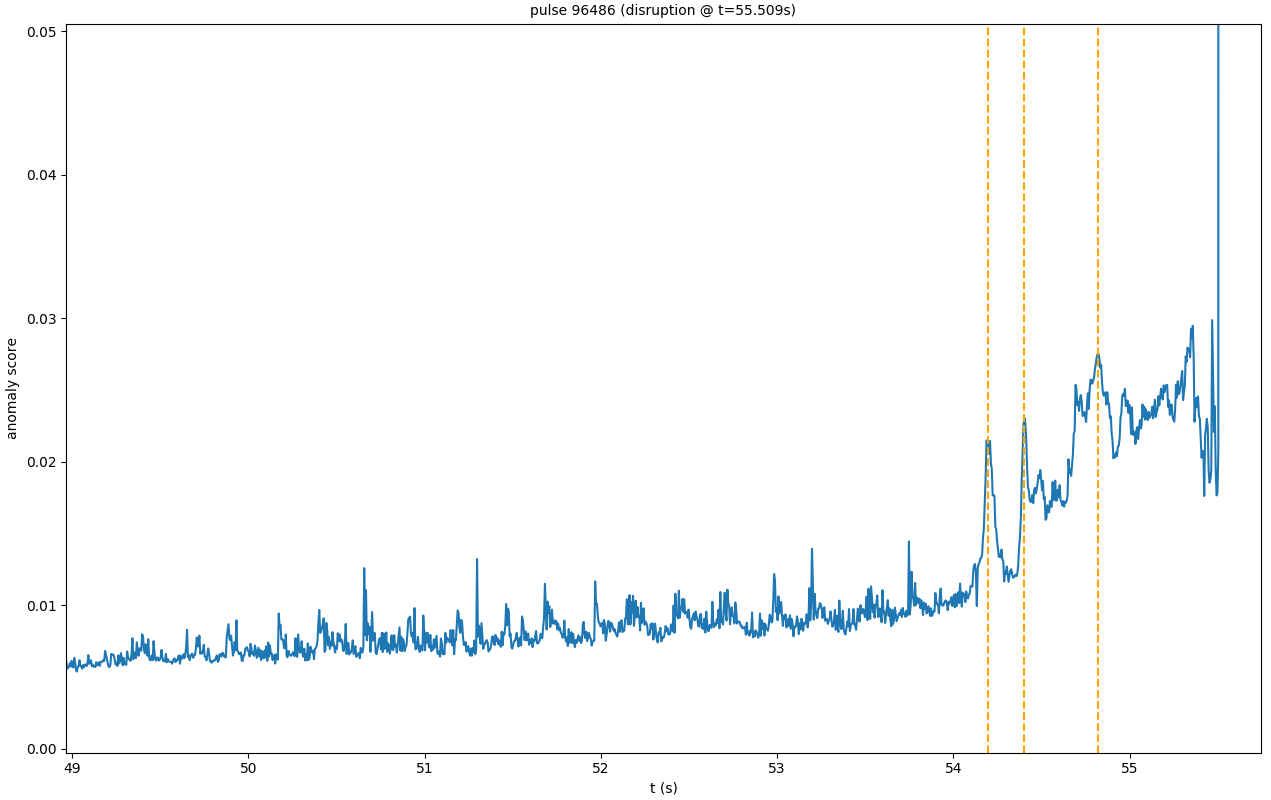}
	\caption{Anomaly score for pulse 96486 with three points of interest depicted in Fig.~\ref{fig_anomaly_tomo}.}
	\label{fig_anomaly_only}
\end{figure}

\begin{figure}[p]
	\centering
	\begin{tabular}{ccc}
		\includegraphics[width=0.32\textwidth]{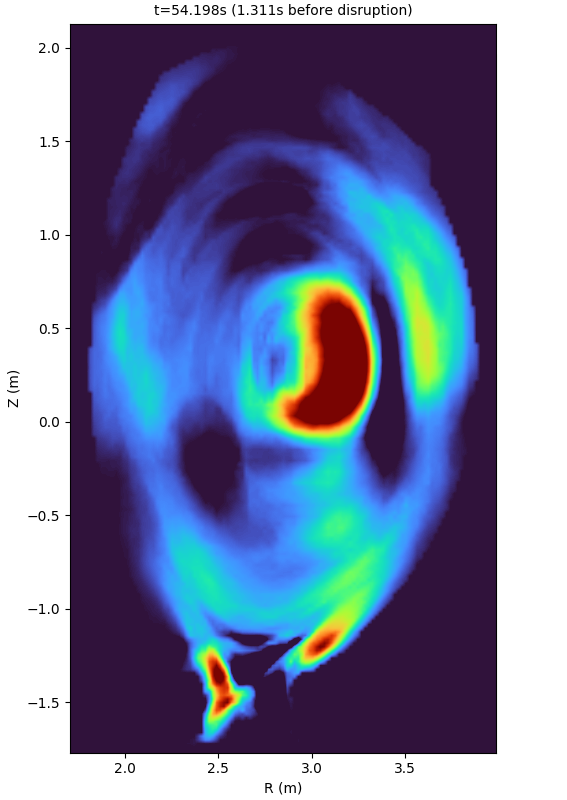} &
		\includegraphics[width=0.32\textwidth]{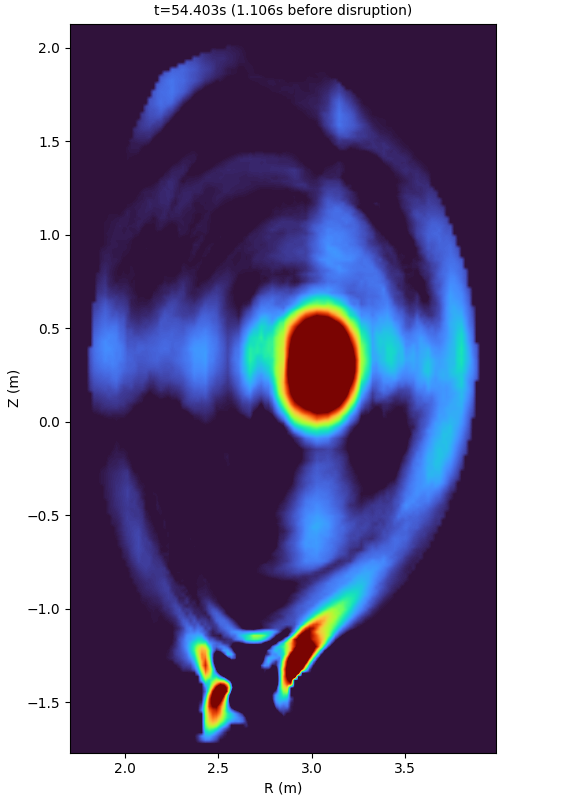} &
		\includegraphics[width=0.32\textwidth]{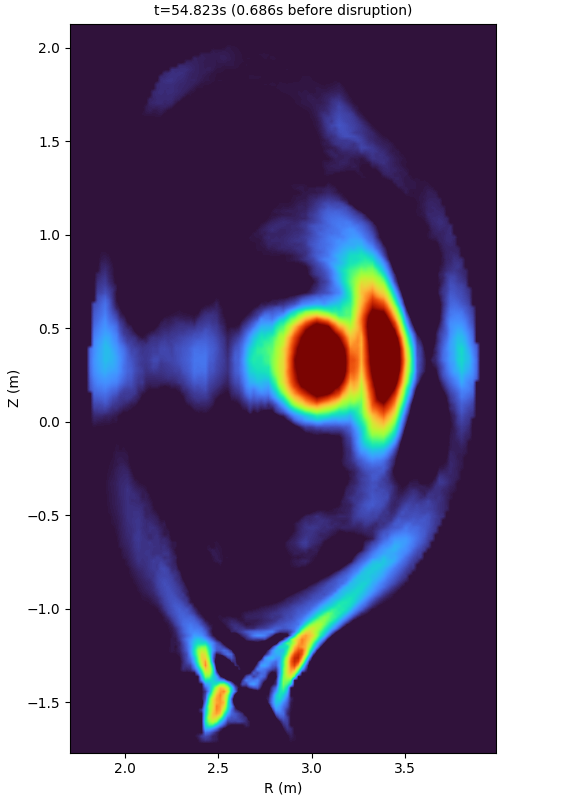} \\
		(\emph{a}) & (\emph{b}) & (\emph{c})
	\end{tabular}
	\caption{Tomographic reconstructions for the three points of interest marked in Fig.~\ref{fig_anomaly_only}.}
	\label{fig_anomaly_tomo}
\end{figure}

Fig.~\ref{fig_anomaly_only} shows the evolution of the anomaly score, where initially it is relatively low, but seems to be on a slightly upward trend through most of the flat-top phase. Then, at about $t$=54.2s, the anomaly score suddenly jumps with the onset of core radiation (Fig.~\ref{fig_anomaly_tomo}\emph{a}), and by $t$=54.4s a core radiation blob is firmly established (Fig.~\ref{fig_anomaly_tomo}\emph{b}).

Subsequently, the core radiation splits into two blobs (Fig.~\ref{fig_anomaly_tomo}\emph{c}), and from $t$=54.7s to $t$=55.4s there are exchanges between the two blobs on a slow time scale, while the anomaly score stays high. When the two blobs finally reunite into single one at $t$=55.4s, a series of MARFE-like instabilities follow in quick succession and the plasma disrupts at $t$=55.5s, even as the core blob was still present.


When cross-checking with other diagnostics, we find that the temperature profile is hollow and the density profile is strongly peaked from $t$=54.2s, the time at which core radiation develops. Then at $t$=55.4s there is an edge cooling consistent with a MARFE, but no locked mode yet. The edge cooling appears to drive MHD activity that eventually results in a locked mode just before the disruption.

Again, different precursors can be used for the purposes of avoidance and mitigation. If the goal is disruption avoidance, then as soon as core radiation develops (as indicated by a rising anomaly score), it should be possible to take preventive action to change the course of events. On the other hand, if disruption mitigation is enough, then an alert on MARFE-like behavior,  which leads to edge cooling and the onset of MHD instabilities such as a locked mode, might be useful if there is still time to react.

One criticism that can be made about this work is that the anomaly score brings nothing new besides what was already available in plain sight. While this is perfectly true, it is precisely in the ability to point out what is hiding in plain sight that the advantages of the proposed approach reside.
Any tools that we can muster to helps us intercept such behavior and bring pulses to safe land will be useful for the operation of JET and future devices, especially ITER.

\section{Conclusion}
\label{sec_conclusion}

In summary, we used deep learning to develop an anomaly detection approach to be applied over the plasma radiation profile, as obtained from bolometer tomography. The approach consists in two different models: one to perform the reconstruction of the radiation profile from bolometer data, and another to detect anomalous profiles by trying to reproduce the behavior of disruptive pulses at each point in time.

For the first model, we could have used a deep neural network as in previous work, but instead we simplified the model by reducing it to a single matrix multiplication layer. This was done in order to be able to incorporate the first model into the second one; at the expense of a slight increase in the reconstruction error, this provides a huge boost in the training performance of both models, enabling full utilization of the available computational resources (GPUs).

For the second model, we used a deep variational autoencoder comprising a series of convolutional layers for encoding, and transposed convolutional layers for decoding the plasma profile. At the heart of this variational encoder is a multivariate normal distribution, whose parameters constitute the latent space which the plasma profile is encoded into and decoded from. The error between the output and the original profile provided as input becomes a measure for the anomaly score of that profile.

Through the analysis of some representative pulses, we have illustrated how the anomaly score can be used as an indicator to identify disruption precursors that are relevant for avoidance and mitigation. In particular, the analysis pinpoints the buildup of core radiation as the earliest sign that the pulse may be heading towards a disruption. This is a recurring pattern in the baseline scenario, where it accounts for at least 50\% of disruptive pulses. There are also edge-related phenomena being highlighted by the anomaly score, but these appear much closer to the disruption, at a time when only mitigation is possible.

The results are consistent with the well-studied phenomena of impurity transport and accumulation, and from this point of view it could be said that this analysis brings nothing new. However, the use of machine learning and anomaly detection on the radiation profile provides a clearer picture of the point in time when problems start to occur, and where in the plasma they occur.

In future work, we plan to extend the approach to other pulses beyond the baseline scenario. However, conducting an analysis across a wider range of pulses will be more computationally expensive, not only in terms of training and test data sizes, but also in terms of the model itself, which should probably become larger in order to accommodate a wider range of physics phenomena.

\section*{Acknowledgments}

\small This work has been carried out within the framework of the EUROfusion Consortium and has received funding from the Euratom research and training programme 2014-2018 and 2019-2020 under grant agreement No 633053. The views and opinions expressed herein do not necessarily reflect those of the European Commission.

\emph{Instituto de Plasmas e Fus\~{a}o Nuclear} (IPFN) received financial support from \emph{Funda\c{c}\~{a}o para a Ci\^{e}ncia e Tecnologia} (FCT) through projects UIDB/50010/2020 and UIDP/50010/2020.

The authors are thankful for the granted use of computational resources provided by the MARCONI-FUSION HPC facility at CINECA, Italy, and by CCFE / UKAEA at Culham, UK.

\bibliographystyle{ans_js}
\bibliography{paper}

\end{document}